\documentclass[final,1p,times]{elsarticle} 

\usepackage{graphicx}
\usepackage{amssymb} 
\usepackage{amsthm} 
\usepackage{lineno}

\journal{Nuclear Physics A} 

\begin{document}

\begin{frontmatter} 

\title{Electroweak Boson-Tagged Jet Event Asymmetries \\ at the Large Hadron Collider}

\author{Ivan Vitev}
\address{Los Alamos National Laboratory, Theoretical Division, Mail Stop B283,  Los Alamos, NM 87544, U.S.A.}


\begin{abstract} 
Tagged jet measurements provide a promising experimental channel to quantify the similarities 
and differences in the mechanisms of jet production in proton-proton and nucleus-nucleus collisions. 
We present the first calculation of the transverse momentum asymmetry of $Z^0/\gamma^*$-tagged jet 
events and the momentum imbalance of $\gamma$-tagged jet events in $\sqrt{s}=2.76$~TeV reactions 
at the LHC. Our results combine the ${\cal O}(G_F\alpha_s^2)$,  ${\cal O}(G_F\alpha_s^2)$ 
perturbative cross sections with the radiative and collisional processes that modify parton showers 
in the presence of dense QCD matter. We find that strong asymmetry momentum and imbalance,
respectively, are generated  in central  Pb+Pb reactions that have little sensitivity to the 
fluctuations of the underlying soft 
hadronic background. We present theoretical model predictions for their shape and magnitude. 
\end{abstract} 

\end{frontmatter} 


\section{Introduction}

Jets tagged by photons ($\gamma$) or electroweak bosons ($W^\pm, \, Z^0$)
are particularly well suited to studying heavy-ion
collisions~\cite{Neufeld:2010fj} since the tagging particle
escapes the region of strongly-interacting matter unscathed.
For example, the CMS collaboration measurements in lead-lead (Pb+Pb)
collisions show absence of significant modification  of
high transverse momentum  $Z^0$ and photon production relative to the
binary collision-scaled proton-proton (p+p) result within the current
statistical and systematic uncertainties.
Thus, in the collinear factorization approach $Z^0$s and $\gamma$s  can provide,
on average, constraints on the energy of the away-side parton shower.
Furthermore, jets tagged by photons or electroweak bosons are largely
unaffected by the fluctuations of the soft hadronic  background
that may complicate the interpretation of di-jet modification in heavy-ion
collisions. By selecting  a suitable range for the transverse momentum of the 
tagging photon ($p_{T_\gamma}$), accessible at both RHIC and the LHC, the 
in-medium modification of parton showers in dense matter created at very different
$\sqrt{s_{NN}}$ can be studied. In the jet suppression region, where
the transverse momentum of the jet $p_{T_{\rm jet}} \geq p_{T_{Z^0, \gamma}}$, the
attenuation of inclusive jets~\cite{He:2011d,:2012is} and photon-tagged 
jets~\cite{Chatrchyan:2012gt,Dai:2012am} can be directly compared.

With this motivation, we present theoretical predictions for the cross section
modification, growth in the transverse momentum asymmetry $A_J$, and  imbalance 
$z_J$ change of $Z^0$-tagged  and photon-tagged
jet events in heavy ion collision and focus on  LHC energies. These variables are
defined as follows:
\begin{equation}
A_J = \frac{ p_{T_{Z^0}} - p_{T_{\rm jet}} } { p_{T_{Z^0}} + p_{T_{\rm jet}} } \, , 
\quad  z_{J\gamma} = \frac{p_{T_{\rm jet}}}{p_{T_\gamma}}  \, .
\end{equation}
Our results combine
the ${\cal O}(\alpha_{\rm em} \alpha_s^2)$, ${\cal O}( G_F \alpha_s^2)$ 
perturbative cross sections with
initial-state cold nuclear matter effects, see for example~\cite{Sharma:2012dy},   and 
final-state parton shower modification and energy dissipation in the QGP.

\section{Asymmetry of $Z^0$-tagged jet events at the LHC}

The calculation of the double differential cross sections for electroweak boson-tagged jet events
 is simpler than the corresponding calculation~\cite{He:2011d} for di-jet 
 events~\cite{Aad:2010bu,Chatrchyan:2011sx}.
The medium-modified tagged cross section per binary scattering is calculated as
follows:
 \begin{eqnarray}
    \frac{1}{\langle N_{bin}\rangle}\frac{d\sigma^{AA}}{dp_{T_{Z^0,\gamma}}dp_{T_{\rm jet}}}
&=& \sum_{q,g} \int_0^1 d\epsilon \frac{P_{q,g}(\epsilon)}{1-[1-f(R)]\epsilon}  R_{q,g} \, 
\frac{d\sigma^{CNM} \left(p_{T_{Z^0,\gamma}},\frac{p_{T_{\rm jet}}}{1-[1-f(R)]\epsilon }\right)  }
{dp_{T_{Z^0,\gamma}}dp_{T_{\rm jet}}} \, .
\label{eq:modify}
\end{eqnarray}
In Eq.~(\ref{eq:modify}) $P_{q,g}(\epsilon)$ is the probability distribution that a
fraction $\epsilon$ of the hard-scattered quark or gluon energy is converted to
a medium-induced parton shower~\cite{Neufeld:2010fj}  and $R_{q,g}$
is the fraction of the corresponding  hard-scattered partons.  Part of the
dependence of the jet cross section on the jet  reconstruction parameters,
such as the radius $R$, is contained  in $d\sigma^{CNM}/dp_{T_\gamma}dp_{T_{\rm jet}}$.
More importantly, the fraction of the parton
shower energy that is simply redistributed inside the jet due to final-state
interactions $f(R)$ also depends on $R$  [$f(R)_{R\rightarrow 0} \rightarrow 0$,  
$f(R)_{R \gg 1} \rightarrow 1$] ~\cite{He:2011d}.
The physics meaning of Eq.~(\ref{eq:modify}) is that the observed $Z^0, \,  \gamma$-tagged
jet cross section in nucleus-nucleus reactions is a probabilistic superposition of cross
sections where the jet is of higher initial transverse momentum
$ p_{T_{\rm jet}}/\{1-[1-f(R)]\epsilon) \}$.

\begin{figure}[!t]
\centerline{
\includegraphics[width = 0.48\linewidth]{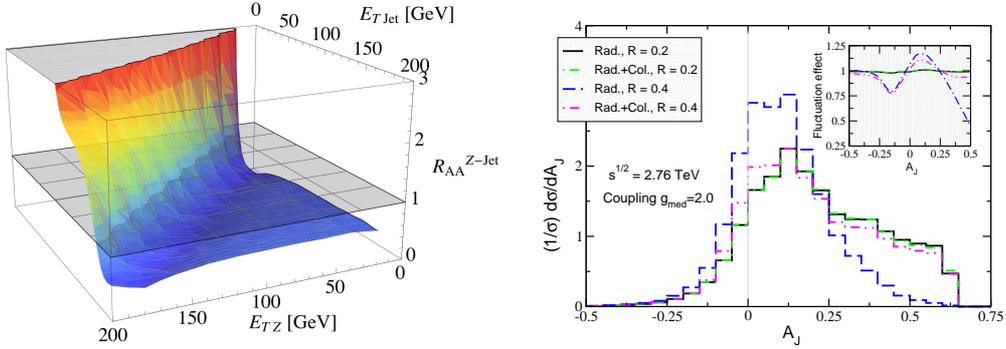} \hspace*{4mm}
\includegraphics[width = 0.45\linewidth]{collision_vary.eps}
}
\caption{Left panel: tagged jet nuclear modification factor, $R_{AA}^{\rm Z^0-jet}$, defined in  
Eq.~(\ref{RAAjet}), including both collisional and radiative energy loss effects. 
Our result is for $R=0.4$ and coupling between the jet and the medium given by 
$g_{\rm med} = 2$, Right panel: $Z^0/\gamma^*$-tagged jet event asymmetry for p+p collisions at  
$\sqrt{s} = 2.76$~TeV for two different $R=0.2, 0.4$. Predictions for central 
Pb+Pb collisions with radiative  medium-induced energy loss are also presented. 
Inset shows the effect of fluctuations in the background subtraction.
}
\label{raa_lhc_tagged}
\end{figure}

Complete information for the cold nuclear matter (CNM) and quark-gluon plasma (QGP) 
effects is contained in the generalized jet nuclear modification factor, $R_{AA}^{\rm Z^0,\gamma-jet}$, 
given by
\begin{eqnarray}
&&R_{AA}^{\rm Z^0,\gamma-jet} ( p_{T_{Z^0,\gamma}},  p_{T_{\rm jet}}; R )   =  
\frac{d\sigma_{AA}}{ d  p_{T_ {Z^0,\gamma}} d p_{T_{\rm jet}} }  \Big/
 \langle  N_{\rm bin}  \rangle
\frac{d\sigma_{pp}}{ d p_{T_{Z^0,\gamma}} d p_{T_{\rm jet}} }  
  \;. \qquad
\label{RAAjet}
\end{eqnarray}
Our simulations include medium-induced parton splitting~\cite{Ovanesyan:2011kn}, here
applied in the soft gluon approximation~\cite{Vitev:2007ve}, and  dissipation of the
medium-induced parton shower energy in the QGP due to collisional processes~\cite{Neufeld:2011yh}.
The nuclear modification factor provides a compact way to quantify 
the effects of the nuclear medium. Our predictions~\cite{Neufeld:2012df} for 
$R_{AA}^{\rm Z-jet} $($R = 0.4$)  
are presented  in the left panel of Fig.~\ref{raa_lhc_tagged}. Since part of the parton shower
energy is redistributed outside of the jet cone radius, the jets are pushed to lower values of 
$p_{T \rm Jet}$. This redistribution results in an enhancement in $R_{AA}^{\rm Z-jet}$ 
in the region of $ p_{T \, \rm Jet} < p_{T \, Z}$ and suppression in  $R_{AA}^{\rm Z-jet}$ 
in the region of $ p_{T \, \rm Jet} > p_{T \, Z}$, which is characteristic of in-medium  tagged-jet 
dynamics~\cite{Neufeld:2010fj}.

Tagged jet asymmetry and imbalance are obtained by changing variables appropriately and
integrating the remaining variable in the specified kinematic domain. For example,  
\begin{eqnarray}
\frac{d\sigma}{dA_J }&=& 
\int_{p_{T\, \rm Jet\, min}}^{p_{T\,\rm Jet \,max}} dp_{T_{\rm jet}}    
\frac{2 p_{T_{\rm jet}}}{(1-A_J)^2}
\frac{d\sigma}{ dp_{T_{Z^0}} dp_{T_{\rm jet}} } \;. \quad
\label{ajcalc}
\end{eqnarray}
In the example that follows  $p_{T\, Z  } \in (80,100)$~GeV and 
$p_{T\, \rm Jet  } > 20$~GeV. in the right panel of Fig.~\ref{raa_lhc_tagged} 
we present the $Z^0/\gamma^*$-tagged jet event asymmetry for central  Pb+Pb collisions 
with radiative (solid and dashed curves)  and radiative+collisional  
(dot-dashed and dot-dot dashed curves)   medium-induced energy 
losses. The collisional energy 
loss has a more pronounced effect in the curve with the larger radius. This occurs 
because collisional energy loss from a parton shower comes primarily from the 
radiated gluons, as demonstrated in~\cite{Neufeld:2011yh}.  With the smaller 
radius most of the gluons are already outside of the jet cone making the extra energy 
loss redundant.  We  point out that background fluctuations again have minimal 
effect when the collisional energy loss is included, as can be checked 
from the insert in Fig.~\ref{raa_lhc_tagged}. The result is a considerable shift in 
$\langle A_J \rangle$ and broadening of the asymmetry distribution.

\section{Momentum imbalance of $\gamma$-tagged jet events at the LHC}

\begin{figure}
\centerline{
\includegraphics[width = 0.48\linewidth]{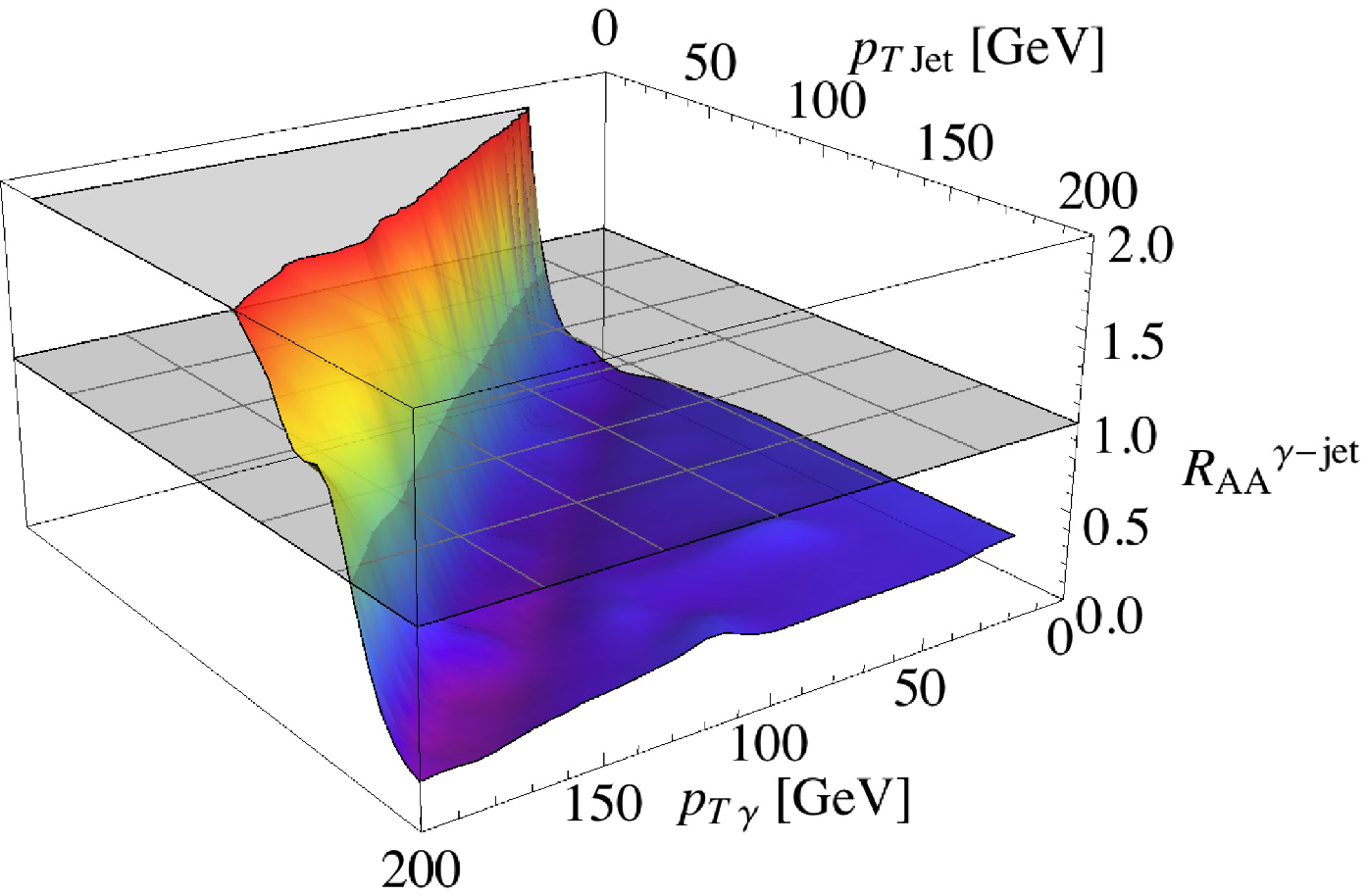} \hspace*{4mm}
\includegraphics[width = 0.45\linewidth]{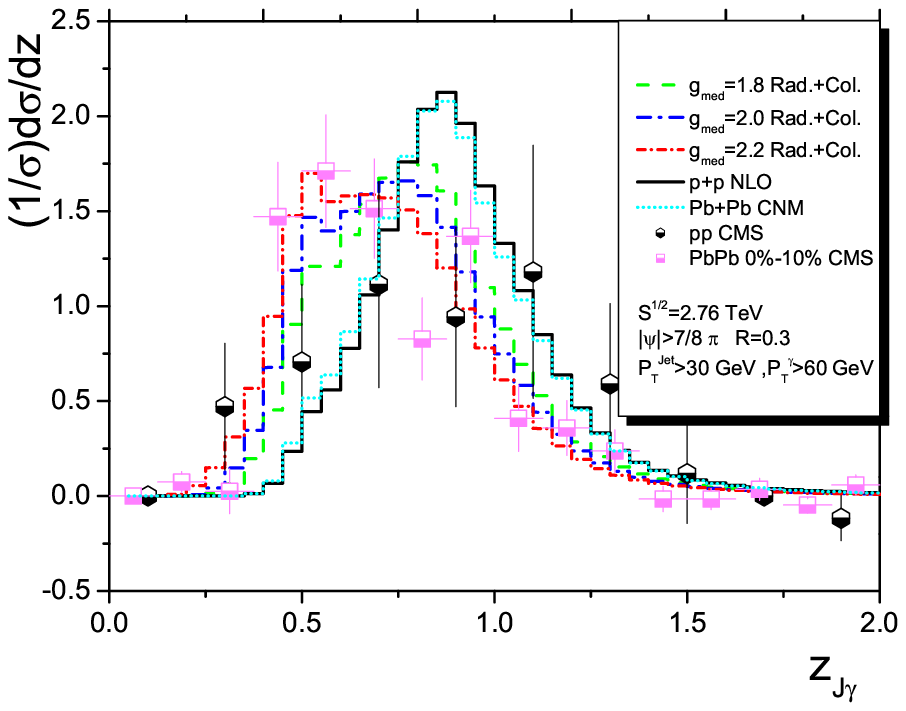}
}
\caption{Left panel: the nuclear modification factor $R_{AA}^{\gamma-\rm jet}$ for the 
isolated photon-tagged jets in the most central Pb+Pb reactions (R=0.3) at $\sqrt{s_{NN}} =2.76$~TeV.  
Right panel: the isolated photon-tagged jet asymmetry distribution for different
coupling  strengths between the jet and the medium. Black and magenta points are CMS
experimental data in p+p and Pb+Pb collision,
respectively.}
\label{raa_lhc_gamma}
\end{figure}

The cross section for isolated photon-tagged jet events is also calculated as
described in  Eq.~(\ref{eq:modify}). The many-body QCD dynamics that modifies 
isolated photon + jet production in
relativistic heavy-ion collisions is manifested in the deviation from the
baseline p+p results, scaled by the number of binary nucleon-nucleon
interactions, and is shown in the left panel of Fig.~\ref{raa_lhc_gamma}.
Our results~\cite{Dai:2012am} are qualitatively similar to
the modification of $Z^0/\gamma^*$ tagged jets~\cite{Neufeld:2012df}. However,
direct comparison between RHIC and LHC is possible for the first time in a
more exclusive channel. 

We can express the $\gamma$-tagged jet event momentum imbalance distribution
as follows:
 \begin{eqnarray}
\frac{d\sigma}{dz_{J}} = \int_{p_{T_{\rm jet}}^{min}}^{p_{T_{\rm jet}}^{max}}d p_{T_{\rm jet}}
\,  \frac{p_{T_{\rm jet}}}{z_{J}^2}
\frac{d\sigma[z_{J},p_{T_\gamma}(z_{J\gamma},p_{T_{\rm jet}})]}{dp_{T_\gamma}dp_{T_{\rm jet}}} \;.
\end{eqnarray}
In our p+p and Pb+Pb calculations at the LHC
we use the CMS experimental cuts~\cite{Chatrchyan:2012gt}  $p_{T_{\rm jet}} > 30$~GeV, $p_{T_\gamma} > 60$~GeV, 
$\vert y^{\gamma} \vert < 1.44$, $\vert y^{\rm jet} \vert < 1.6$,
$\vert \phi^{\rm jet}-\phi^\gamma \vert > \frac{7}{8}\pi$. We implement a $k_T$ algorithm with a 
radius parameter $R = 0.3$ for the jet and isolation criterion that requires
the total energy within a cone of radius $R_{\rm iso.}= 0.4$ surrounding the photon 
direction to be less than 5~GeV.
The normalized momentum imbalance distribution $({1}/{\sigma})
{d\sigma}/{dz_{J}}$ is given in the right panel of Fig.~\ref{raa_lhc_gamma}. The solid black
line shows the p+p calculation and the circles represent the CMS result with
large error bars~\cite{Chatrchyan:2012gt}. The dotted cyan line includes  cold  nuclear matter effects
in central Pb+Pb  reactions. These CNM  effects do not affect the $z_{J}$
distribution appreciably. The physics responsible for the difference between
p+p and A+A reactions is then contained in the final-state QGP-induced parton
splitting and the dissipation of the parton shower energy in the medium. The
parameter that controls the strength of the coupling between the jet constituents
and the strongly interacting matter is $g_{\rm med}$. We investigate a
range  of values  $ g_{\rm med} = 1.8 \ ({\rm green \ dashed} )$,  $2.0 \
({\rm blue \  dot-dashed} )$, $2.2 \ ({\rm red \ short  \ dot-dashed} ) $
that has worked well in describing  the di-jet asymmetry distribution and
in predicting the inclusive jet suppression at the LHC~\cite{He:2011d}.
The same range of coupling strengths has been used to predict the asymmetry
distribution of $Z^0$+jet events in heavy-ion collisions~\cite{Neufeld:2012df}.

\section{Summary}

We presented  selected results from the first studies~\cite{Dai:2012am,Neufeld:2012df}  
of the nuclear modification of  $Z^0, \gamma$-tagged jet cross sections and the corresponding 
changes in their  transverse momentum asymmetry and momentum imbalance distributions.  We found 
that a pQCD approach~\cite{He:2011d} that describes well the inclusive jet  suppression~\cite{:2012is} 
and enhancement of the di-jet  asymmetry~\cite{Aad:2010bu,Chatrchyan:2011sx}  in heavy ion collisions at the LHC
can predict the shape modifications ($A_J$, $z_J$)  and cross section attenuation   of  the electroweak boson-tagged jet events recently measured by the 
ATLAS~\cite{atlas} and CMS~\cite{Chatrchyan:2012gt} collaborations in $\sqrt{s}=2.76$~TeV Pb+Pb collisions.

\end{document}